# Is High-density Amorphous Ice Simply a 'Derailed' State along the Ice I to Ice IV Pathway?

*Jacob J. Shephard,[a] Sanliang Ling,[a] Gabriele C. Sosso,[b] Angelos Michaelides,[b] Ben Slater[a] and Christoph G. Salzmann*[a]*

[a] Department of Chemistry, University College London, 20 Gordon Street, London WC1H 0AJ, United Kingdom.

[b] Thomas Young Centre, Department of Physics and Astronomy, and London Centre for Nanotechnology, University College London, Gower Street, London WC1E 6BT, United Kingdom.

**Corresponding Author**

* c.salzmann@ucl.ac.uk

**ABSTRACT**

The structural nature of high-density amorphous ice (HDA), which forms through low-temperature pressure-induced amorphization of the 'ordinary' ice I, is heavily debated. Clarifying this question is not only important for understanding the complex condensed states of $H_2O$ but also in the wider context of pressure-induced amorphization processes, which are encountered across the entire materials spectrum. We first show that ammonium fluoride ($NH_4F$), which has a similar hydrogen-bonded network to ice I, also undergoes a pressure collapse upon compression at 77 K. However, the product material is not amorphous but $NH_4F$ II, a high-pressure phase isostructural with ice IV. This collapse can be rationalized in terms of a highly effective mechanism. In the case of ice I, the orientational disorder of the water molecules leads to a deviation from this mechanism and we therefore classify HDA as a 'derailed' state along the ice I to ice IV pathway.

A simple compression experiment of the 'ordinary' ice I$h$ at liquid-nitrogen temperature has left a remarkable legacy and tremendous impact across the materials sciences.[1] It marked the discovery of high-density amorphous ice (HDA) which provided the first example for the process of pressure-induced amorphization (PIA). Countless other examples of PIA have been reported since for a wide range of inorganic, organic and even biochemical materials.[2] However, questions related to the detailed mechanisms of the PIA processes are still under debate. In particular since some reported cases of PIA were later attributed to either the formation of nanocrystals, chemical decomposition or the non-hydrostatic compression conditions.[2-7]

The PIA of ice I$h$ was first rationalized in terms of a thermodynamic melting process implying that HDA is a glassy state of high-pressure water.[1, 8-12] This, together with the transformation of HDA to low-density amorphous ice upon decompression or heating at ambient pressure[10] provided an experimental background for the two-liquid model of water and the 2$^{nd}$ critical-point scenario.[13-15] Yet, later, it was also argued that HDA forms due to a mechanical instability and can therefore be understood as a collapsed 'ill-crystalline' material that lacks a connection with the liquid phase.[16-23] A temperature-driven cross-over between these two scenarios has also been suggested.[9, 17, 20] Finally, it was even speculated that HDA may consist of strained microdomains of crystalline high-pressure phases of ice.[24-25] In summary, it is safe to state that the structural and thermodynamic nature of HDA is still poorly understood. Due to the pivotal position of HDA not only for understanding the condensed states of $H_2O$ but also in the wider context of PIA processes, a deeper knowledge of the mechanism of the PIA of ice I is of paramount importance.

Here we first investigate the low-temperature compression behavior of hexagonal and stacking-disordered ammonium fluoride ($NH_4F$) whose hydrogen-bonded networks are similar compared to the corresponding members of the ice I family. From this, we then aim to gain new insights into the origin of the pressure collapse of ice upon low-temperature compression and to propose a structural mechanism for the PIA of ice I which we also test with density functional theory (DFT) calculations.

Ice I and $NH_4F$ I are similar materials from the structural point of view. Figure 1a shows the isostructural hydrogen-bonded networks of the hexagonal ice I$h$ and $NH_4F$ I$h$ materials which are the stable phases at ambient conditions. The hydrogen bonds in $NH_4F$ I$h$ are less than 2% shorter than in ice I$h$.[26] A metastable variant of ice I$h$ exists which contains interlaced sequences of cubic and hexagonal stacking, generally known as stacking disordered ice (ice I$sd$).[27-30] The most cubic ice I$sd$ prepared so far, which will be used in the following, was obtained by slowly heating the ice II high-pressure phase at ambient pressure.[30] The material obtained by heating the $NH_4F$ II high-pressure phase at ambient pressure has been labelled as $NH_4F$ V.[31] Yet, as we will show, this material is, in analogy with the situation for ice, best described as stacking-disordered ammonium fluoride I ($NH_4F$ I$sd$).

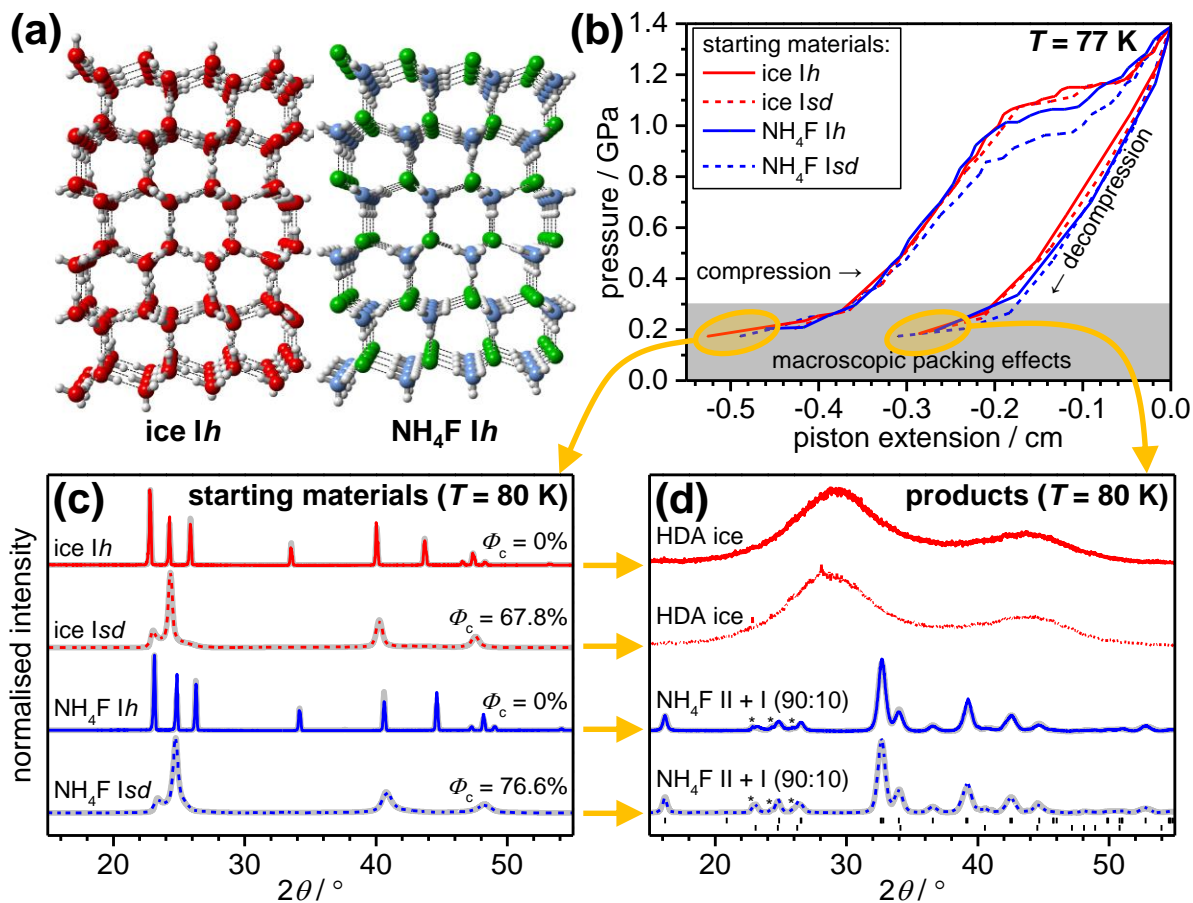

**Figure 1.** Ice I and NH$_4$F I show remarkably similar pressure collapses upon compression at 77 K. (a) Crystal structures of ice I$h$ and NH$_4$F I$h$. (b) Changes in sample volume, as indicated by the piston extension, upon compression and decompression at 77 K of equal amounts of ice I and NH$_4$F I materials ($n(H_2O) = n(NH_4^+) + n(F^-)$). Powder X-ray diffraction patterns of the materials (c) before and (d) after compression. The percentages of cubic stacking, $\Phi_c$, of the starting materials are given in (c) and the tickmarks in (d) indicate the positions of the Bragg peaks of NH$_4$F II and I$h$, respectively. Thick grey lines in (c) and (d) are Rietveld or MCDIFFaX[29, 31] fits.

Remarkably, as shown in Figure 1b, NH$_4$F I$h$ and NH$_4$F I$sd$ show similar collapses upon compression at 77 K in terms of the onset pressures and overall volume changes compared to the ice I samples. The X-ray diffraction patterns of the starting materials are shown in Figure 1c. Using our MCDIFFaX software,[30, 32] the ice I$sd$ and NH$_4$F I$sd$ samples were shown to contain 67.8% and 76.6% cubic stacking, respectively. Figure 1d shows that the diffraction patterns obtained after the compression of the ice materials are consistent with HDA in line with previous studies.[1, 33]

On the basis of the similar pressure collapses, we initially assumed that the NH$_4$F samples had also undergone transitions to high-density amorphous materials in analogy to what has been observed for ice. Surprisingly, the diffraction patterns of the NH$_4$F materials after compression shown in Figure 1d revealed that they actually consist of crystalline mixtures of ~90% NH$_4$F II and ~10% NH$_4$F I. Upon comparison of the crystal structure of NH$_4$F II[34] with the known phases of ice it was realized that NH$_4$F II is isostructrual with the metastable ice IV.[35] Although NH$_4$F II is the first high-pressure phase to form upon compression of NH$_4$F I$h$ at room temperature, it is probably not the stable phase at 77 K and ~1 GPa due to the pronounced slope of the NH$_4$F II / III phase boundary.[36] This suggests that the NH$_4$F II, which results from the low-temperature compression, is a kinetic product that forms as a consequence of a favorable mechanistic pathway.

The transformation of NH$_4$F I to NH$_4$F II can be understood on the basis of a remarkable mechanism that achieves a 37% increase in density while only breaking one in four hydrogen bonds. This collapse was first described by Engelhardt and Kamb for the transition from the hypothetical fully cubic ice I (ice I$c$) to ice IV.[35] The corresponding collapse starting from the hexagonal starting material was later described in the NH$_4$F literature.[34] The mechanistic details of this collapse, which we refer to as the Engelhard-Kamb collapse (EKC), are shown in Figure 2. The ice I$c$/I$h$ networks contain identical layers consisting of puckered six-membered rings in the armchair conformation. These networks are shown in Figure 2a with the network nodes of the individual layers highlighted in different colors.

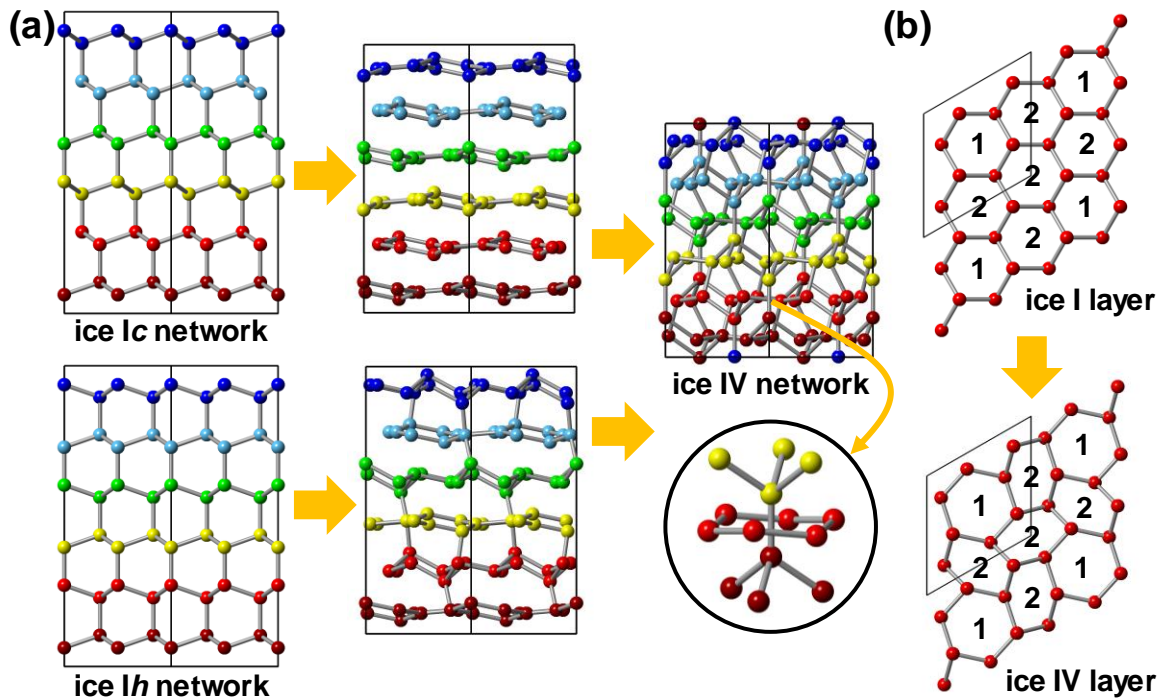

**Figure 2.** Formal mechanism of the Engelhard-Kamb collapse.[34-35] (a) Transformation of 2×2×3 supercells of the ice/$NH_4F$ I$h$ unit cells viewed along [100] to the ice IV/$NH_4F$ II unit cell in the hexagonal setting. The spheres indicate the nodes of the hydrogen-bonded networks and are colored differently for each layer. The 'threading-through' structural feature is shown in the inset. (b) Transformation of a single layer of the I$h$/I$c$ networks to the ice IV network viewed along [001].

The transition from the ice I$c$ to the ice IV network requires breaking of the inter-layer hydrogen bonds followed by flattening and ultimately a complete 're-buckling' of the layers.[35] The last step is the formation of hydrogen bonds right through the center of the six-membered rings of the layers above and below towards water molecules of the second layers above and below.

This 'threading-through' is the hallmark structural feature of the ice IV network which 'pulls' the layers closer together and thereby achieves the increase in density (inset in Figure 2a). Starting off from the ice I$h$ network, the 're-buckling' of layers is not required. Instead, the individual layers need to shift or rotate with respect to one another to achieve the 'threading-through' of the six-membered rings. Since both the ice I$c$ and I$h$ networks can undergo the EKC, it is of course also possible to start from stacking disordered starting materials.

The effect of the ice IV network formation on an individual layer is shown in Figure 2b. The six-membered rings labelled with '1' are the ones that experience 'threading-through' with hydrogen bonds during the transition to the ice IV network. This leads to a flatting of these rings, a slight increase in diameter and rotation. These changes in the '1' rings distort the rings labelled with '2' which either raises or lowers the nodes which are not members of the '1' rings so that they can form the hydrogen bonds with two layers above or below.

A remarkable feature of the EKC is that the density increase is highly anisotropic and achieved almost exclusively by contraction along [001].[34] It is difficult to imagine that there is another mechanism that achieves a similar increase in density while breaking fewer hydrogen bonds. So, the EKC seems to be a particularly important mechanism for densification under kinetically controlled low-temperature conditions.

The similar onset-pressures and volume changes of the low-temperatures collapses of ice I and NH$_4$F I suggest that both materials suffer from a similar mechanical instability that marks the onset of the EKC. However, in both cases, the EKC is not followed through entirely at 77 K. In the case of NH$_4$F I, the conversion to NH$_4$F II is ~90 w% leaving ~10 w% of NH$_4$F I unconverted. The incomplete conversion is attributed to the build-up of macroscopic strain environments which arise from the highly anisotropic nature of the EKC.

In the case of ice, it is important to recall that the ice I materials are hydrogen-disordered which means that they display disorder with respect to the orientations of the hydrogen-bonded molecules (Figure 1a). As mentioned earlier, during the EKC, the interlayer hydrogen bonds are broken and reformed with water molecules one and two layers above or below. The chances of successfully connecting a broken hydrogen bond to another water molecule are 50% since water molecules can be either hydrogen-bond donors or acceptors. This implies, that in case of ice I, not all broken hydrogen bonds will necessarily thread-through the six-membered rings in the final stage of the EKC but form hydrogen-bonds with other close water molecules instead that fulfil the donor / acceptor conditions. The resulting material is HDA which displays a lack of long-range order and a density somewhat lower than ice IV. Accordingly, HDA can be classified as a 'derailed' state along the EKC and we provide further arguments for this in the following.

The hydrogen-bond donor / acceptor mismatch in ice could in principle be resolved with molecular reorientations. These are known to be very slow at 77 K,[37] but have been shown to 'unfreeze' during the glass transition of HDA at higher temperatures.[38] When HDA is heated at around its 'natural' pressure of 1 GPa, the amorphous sample becomes denser at first followed by crystallization.[39-40] The crystallization to ice IV has been shown to have the lowest activation energy out of all the possible crystallization pathways around 1 GPa.[41-43] This means that the 'derailed' state of HDA can be brought 'back on track' with respect to the EKC by thermal annealing under pressure as dynamic molecular reorientations resolve the problem of the hydrogen-bond donor / acceptor mismatches. The similar structures of pressure-annealed HDA and ice IV are also evident from their Raman spectra.[40, 44] Furthermore, the first strong diffraction peak of pressure-annealed HDA, which reflects the intermediate range structural order, is in the same position as the strongest Bragg peaks of ice IV.[39-41]

An interesting question now arises if the 'derailment' upon low-temperature compression of ice I could be prevented by using hydrogen-ordered starting materials. The hydrogen-ordered ice I$c$ (ice XI$c$)[45] is a promising starting material since the EKC from a cubic starting material does not require the translational movement of layers. All inter-layer hydrogen bonds point in the same direction in ice XI$c$ which means that the problem of the hydrogen-bond donor / acceptor mismatches does not apply. Unfortunately, ice I$c$ has not been prepared so far,[30] and consequently it is unknown if and how ice XI$c$ can be prepared. Also, the fact that ice I$h$ becomes only partially hydrogen-ordered upon doping with potassium hydroxide below 77 K[46-48] illustrates how much of a challenge the low-temperature compression of ice XI$c$ would be.

To investigate the feasibility of the ice XI$c$ to ice IV pathway, we therefore performed a transition-state search using the solid-state nudged-elastic-band (NEB) method[49] at the dispersion-corrected DFT level. The solid-state NEB method couples the atomic and lattice degrees of freedom of a crystal structure, and it requires a reasonable initial guess of the transition pathway from XI$c$ to ice IV. A minimum-energy contiguous path was obtained after optimization, and the energy profile and cell volumes of all frames on the transition path are shown in Figure 3. The activation energy for the phase transition is found to be 11 kJ/mol, about twice as large as the enthalpy difference between ice IV and XI$c$. The initial configurations are associated with the formation of point defects that resemble interstitials. The volume changed slowly during the first stages of the transition but close to and after the transition state, there was a marked drop. The largest change in volume at the very end of the transition is clearly associated with the formation of hydrogen bonds passing through the six-membered rings (Figure 2). We note that our estimated activation energy is about an order of magnitude higher than what is expected for a purely enthalpy-motivated transition at 77 K. Several factors could contribute to this overestimation. First, we started from a perfect lattice of ice XI$c$. In ice, non-negligible concentrations of kinetically trapped point and line defects will be present, which, together with grain-boundary effects, may act as the seeds for the transition to ice IV, effectively lowering the transition barrier. Second, our predicted transition mechanism only represents one of the possible transition pathways; lower-energy transition mechanisms may exist for larger simulation cells, where the point defects, that facilitate the transition, are more spatially separated, allowing for greater lattice relaxation. Third, overestimation of the activation barrier may also be enhanced from the DFT method we use (PBE+D3) which 'overbinds' the ice phases.[50] Nevertheless, overall it seems very plausible that ice XI$c$ is likely to undergo the EKC upon low-temperature compression and to avoid the 'derailment' to HDA.

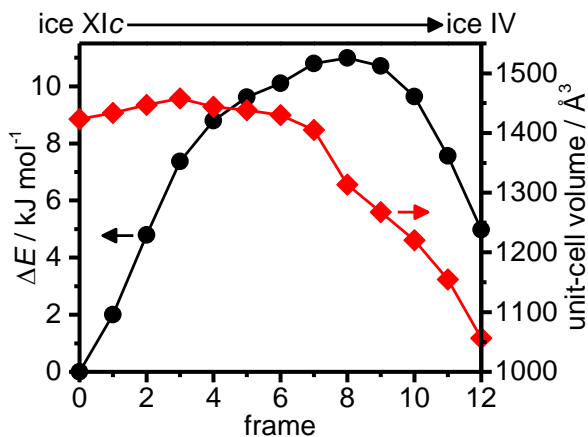

**Figure 3.** Computational 'feasibility test' of the ice XI$c$ to ice IV transition using solid-state nudged-elastic-band DFT calculations.

In conclusion, we have shown that the hydrogen-bonded networks of ice I and NH$_4$F I both suffer from similar mechanical instabilities upon low-temperature compression which mark the onsets of EKCs in these materials. Yet, both materials do not follow the EKC through completely. This is most likely due to the build-up of macroscopic strain in NH$_4$F and the hydrogen-bond donor / acceptor mismatches in ice as illustrated by the DFT calculations of the ice XI$c$ to ice IV pathway. Put simply, the low-temperature PIA of ice I is a consequence of its hydrogen disorder. This means that HDA does not represent the glassy high-pressure liquid, at least not the material that is obtained from the low-temperature compression of ice I at 77 K. Furthermore, this study also showcases the high efficiency of the EKC in achieving large density increases under kinetically controlled conditions. Future work should focus on the low-temperature compressions of other structurally related materials such as diamond,[32] silicon[51] and the large family of AB materials including silver iodide, zinc sulfide, gallium phosphide and silicon carbide.

**Experimental and Computational Methods**

NH$_4$F I$h$ (99.99%) was purchased from Sigma Aldrich and stored under dry nitrogen. For the compression experiments at 77 K, 404.0 mg of NH$_4$F I$h$ were placed inside an indium cup which was quickly transferred into a stainless-steel piston cylinder precooled with liquid nitrogen. The ice I$h$ samples were prepared by pipetting 393.0 mg of MilliQ water into an indium cup inside a precooled piston cylinder. The compression to 1.4 GPa at 77 K was performed with a 30-tonne hydraulic press. Changes in sample volume were recorded using a GT5000RA-L25 positional transducer from RDP electronics. To obtain NH$_4$F I$sd$, NH$_4$F II was prepared in a first step by compressing 404.0 mg of $NH_4F$ I$h$ to 0.25 GPa at room temperature, followed by cooling to 77 K under pressure, and heating from 77 K to 203 K at ambient pressure. Ice I$sd$ was obtained by heating an ice II sample from 77 K to 190 K at ambient pressure. Both stacking-disordered materials were subjected to the same compression experiments at 77 K as previously described for the hexagonal materials. After the compression experiments, the samples were recovered under liquid nitrogen at ambient pressure, and characterized with X-ray powder diffraction using a custom-made sample holder with Kapton windows mounted on a Stoe Stadi-P diffractometer (Ge 111 monochromated Cu K$_\alpha$, 40 kV, 30 mA). Data were collected using a Mythen 1K linear detector and the temperature of the sample was controlled with an Oxford Instruments CryojetHT.

The solid-state NEB calculations were performed using the atomic simulation environment[52] and the TSASE code.[49] Thirteen frames of the 2x2x3 supercells from ice XI$c$ to ice IV were considered on the transition path. The transition-state search was considered converged if the maximum atomic force falls below 0.1 eV/Å. The energies, forces and stress tensors were calculated using the CP2K code, which uses a mixed Gaussian/planewave basis set.[53-54] We employed double-ζ polarization quality Gaussian basis sets and a 600 Ry plane-wave cutoff for the auxiliary grid, in conjunction with the Goedecker-Teter-Hutter pseudopotentials.[55-56] All calculations were performed using the popular pairwise-additive descriptions of the dispersion interactions, *i.e.* the D3 method with the Axilrod-Teller-Muto three-body terms,[57] in combination with the PBE functional.[58]

ASSOCIATED CONTENT

## AUTHOR INFORMATION

**Notes**

The authors declare no competing financial interests.

## ACKNOWLEDGMENT

We thank M. Vickers for help with the X-ray measurements, J. K. Cockcroft for access to the cryojet, and P. F. McMillan and J. S. Tse for helpful discussions. Funding was received from the Royal Society (C.G.S., UF150665 University Research Fellowship & A.M., Wolfson Research Merit Award), the Leverhulme Trust (J.J.S. & C.G.S., RPG-2014-04), EPSRC (S.L. & B.S., EP/K039296/1) and the European Research Council under the European Union's Seventh Framework Programme (G.C.S & A.M., FP/2007-2013 and ERC Grant Agreement No. 616121 – HeteroIce project). We also acknowledge the use of Emerald, a GPU-accelerated High Performance Computer made available by the Science & Engineering South Consortium operated in partnership with the STFC Rutherford-Appleton Laboratory and use of the ARCHER UK National Supercomputing Service (http://www.archer.ac.uk) through the Materials Chemistry Consortium via the EPSRC Grant No. EP/L000202.